\documentclass[conference]{IEEEtran}
\IEEEoverridecommandlockouts

\usepackage[nolist,nohyperlinks]{acronym}
\def\BibTeX{{\rm B\kern-.05em{\sc i\kern-.025em b}\kern-.08em
    T\kern-.1667em\lower.7ex\hbox{E}\kern-.125emX}}

\usepackage{enumerate}
\usepackage{algorithm2e}
\usepackage{algpseudocode}
\usepackage{graphics}
\usepackage{xparse}
\usepackage{csvsimple}
\usepackage{balance}
\usepackage{ifthen}

\usepackage{hyperref}

\usepackage{cite}
\usepackage{amsmath,amssymb,amsfonts}
\usepackage{graphicx}
\usepackage{textcomp}
\usepackage{xcolor}
\usepackage{soul}
\usepackage{listings}
\usepackage{multirow}
\usepackage[export]{adjustbox}
\usepackage{mathabx}
\usepackage{array}
\usepackage{mdwmath}
\usepackage{mdwtab}
\usepackage{eqparbox}
\usepackage{url}
\usepackage{color}

\newcommand{\version}{final}

\newcommand{\LEARNING}{\emph{Learning Plane}\xspace}
\newcommand{\PREDICTION}{\emph{Prediction Plane}\xspace}
\newcommand{\OPERATION}{\emph{Operational Plane}\xspace}

\usepackage{xparse}

\ExplSyntaxOn
\NewExpandableDocumentCommand{\ifstringsequalTF}{mmmm}
 {
  \str_if_eq:eeTF { #1 } { #2 } { #3 } { #4 }
 }
\NewExpandableDocumentCommand{\stringcase}{mO{}m}
 {
  \str_case_e:nnF { #1 } { #3 } { #2 }
 }
\ExplSyntaxOff

\ifstringsequalTF{\version}{draft}
  {
    \newcommand{\added}[1]{{\color{blue}#1}}
    \newcommand{\deleted}[1]{{\color{red}\st{#1}}}
    \newcommand{\noted}[1]{{\color{orange}#1}}
  }
  {
    \ifstringsequalTF{\version}{final} {
        \newcommand{\added}[1]{#1}
        \newcommand{\deleted}[1]{}
        \newcommand{\noted}[1]{}
    } 
    {
      \ifstringsequalTF{\version}{old} {
        \newcommand{\added}[1]{}
        \newcommand{\deleted}[1]{{\color{red}#1}}
        \newcommand{\noted}[1]{}
      }
      {
        \newcommand{\added}[1]{{#1}}
        \newcommand{\deleted}[1]{}
        \newcommand{\noted}[1]{}
      }
    }    
  }

\begin{document}
\def\thetitle{Machine Learning-based Early Attack Detection Using Open RAN Intelligent Controller}
\title{\thetitle}

\author{
\IEEEauthorblockN{
    Bruno Missi Xavier\IEEEauthorrefmark{2}\IEEEauthorrefmark{3},
    Merim Dzaferagic\IEEEauthorrefmark{3},
    Diarmuid Collins\IEEEauthorrefmark{3},
    Giovanni Comarela\IEEEauthorrefmark{2}, \\
    Magnos Martinello\IEEEauthorrefmark{2} and
    Marco Ruffini\IEEEauthorrefmark{3}
}
\IEEEauthorblockA{\IEEEauthorrefmark{2}
    Federal University of Espírito Santo, Espírito Santo, Brazil}
\IEEEauthorblockA{\IEEEauthorrefmark{3}
    Trinity College Dublin, Ireland}
\IEEEauthorblockA{Email: bruno.xavier@ifes.edu.br}
}
\maketitle
\noted{BRUNO: We should change the title. I suggested some keywords, however, we should review it.}

\noted{Suggestions for technical symposia:}

\noted{ - Mobile and Wireless Networks}

\noted{ - Machine Learning for Communications}

\begin{abstract}
We design and demonstrate a method for early detection of Denial-of-Service attacks. The proposed approach takes advantage of the OpenRAN framework to collect measurements from the air interface (for attack detection) and to dynamically control the operation of the Radio Access Network (RAN). For that purpose, we developed our near-Real Time (RT) RAN Intelligent Controller (RIC) interface. We apply and analyze a wide range of Machine Learning algorithms to data traffic analysis that satisfy the accuracy and latency requirements set by the near-RT RIC. Our results show that the proposed framework is able to correctly classify genuine vs. malicious traffic with high accuracy (i.e., $95\%$) in a realistic testbed environment, allowing us to detect attacks already at the Distributed Unit (DU), before malicious traffic even enters the Centralized Unit (CU). 
\end{abstract}

\begin{IEEEkeywords}
\added{machine learning; mobile networks; 4G and 5G}
\vspace{-1.5em}
\end{IEEEkeywords}

\section{Introduction}\label{sec:intro}

The growing number of services that run on top of cellular networks pose new challenges in ensuring service availability. The common approach to service availability improvement involves network planning, resource allocation optimization, and network densification. However, the source of service outages is often not related to the network configuration but originates from different types of malicious service attacks \cite{eliyan2021and, zoure2022network, wang2021machine, rose2021intrusion}. These security threats affect user satisfaction and result in financial losses.





As highlighted by the authors of \cite{bilogrevic2010security}, security and privacy in cellular networks can be achieved at the air interface, the operator's internal network, and the inter-operator links. Depending on the type of attack (e.g., passive in which the attacker only listens to the traffic, active in which traffic is being modified or injected into the communication), detecting the threat can be a challenge. Passive attacks often result in privacy breaches \cite{nasr2019comprehensive}, while active attacks result in service disruptions \cite{khan2021security}. The focus of our work is on the early detection of active attacks in cellular networks. 

\deleted{The authors of \mbox{\cite{novaes2021adversarial}} emphasize the importance of protecting network services from \mbox{\ac{ddos}} attacks. The authors highlight the vulnerability of the centralized control plane in the \mbox{\ac{sdn}} architecture. Current solutions to the problem heavily depend on the use of \mbox{\ac{ml}} techniques to identify attacks in the network and to activate mitigation procedures. The solutions include the use of classification methods to implement the congestion control mechanisms \mbox{\cite{jay2019deep,wei2021congestion}}, to identify the traffic inside the network \mbox{\cite{xiong2019switches,xavier2021programmable,paolucci2021nns,lee2020switchtree,barradasflowlens}}, to detect \mbox{\cite{meng2019towards,wang2019constructing}} and mitigate the volumetric \mbox{\ac{ddos}} attack \mbox{\cite{ding2021volumetricddos,zhang2020poseidon,liu2021jaqen,shaik2019new}}. In mobile networks, \mbox{\ac{ml}} is used to support the authentication \mbox{\cite{liu2022online}} procedures, and to identify attacks \mbox{\cite{li2021physical,tokgoz2022physical}} in \mbox{\ac{mimo}} systems.}

\added{Recent publications have taken advantage of the processing power and programmability available in the new generation of switches to introduce new \ac{ids} and \ac{dpi} paradigms. Both approaches heavily rely on the use of \ac{ml} techniques to allow the identification of the traffic \cite{xavier2022tnsm, paolucci2021nns, lee2020switchtree, barradasflowlens}, to detect and to mitigate the volumetric \ac{ddos} attacks \cite{ding2021volumetricddos, zhang2020poseidon, liu2021jaqen} inside communication networks. In cellular networks, it is very important to identify the malicious flow as early as possible to prevent it from reaching the \ac{sdn} architecture \cite{novaes2021adversarial} or interrupting services on the edge of the network (e.g., mobile base station).}

\deleted{In \mbox{\cite{ramanathan2018senss}}, the authors explain that the volumetric \mbox{\ac{ddos}} attacks cannot be handled by the victim alone, and require help from the network. An important step in the problem resolution is early detection and localization of the source of the attack. This approach allows us to cut off the malicious traffic before it spreads throughout the network. In terms of cellular networks, the traditional approach to such security threats would involve deep packet inspection in the core network (i.e., on switches and firewalls). However, as highlighted by the authors of \mbox{\cite{bonati2021intelligence}}, the fifth (5G) and sixth (6G) generations of cellular networks will move from inflexible and monolithic networks to agile and disaggregated architectures based on softwarization and virtualization. Additionally, they stress that the OpenRAN Alliance introduces openness and, more importantly, intelligence and threat assessment to the very edge of the network. The added intelligence adds advanced radio resource management, self-organization, and zero-touch optimization to the \mbox{\ac{ran}}. The research in this area mainly focuses on optimizing radio resources, and the dynamic reconfiguration of network entities to match the communication requirements \mbox{\cite{bonati2021intelligence,polese2022understanding}}. In \mbox{\cite{polese2022understanding}}, the authors argue that the security implications of the OpenRAN architecture have to be studied and understood to make it a future-proof alternative to traditional \mbox{\ac{ran}} deployments. }

\added{As highlighted by the authors of \mbox{\cite{ramanathan2018senss}}, the volumetric \mbox{\ac{ddos}} attacks cannot be handled by the victim alone, and require help from the network. In terms of cellular networks, the traditional approach relies on the core network to deal with the malicious flows \cite{iavich2021novel}. However, the fifth (5G) and sixth (6G) generations of cellular networks will move from inflexible and monolithic networks to agile and disaggregated architectures based on softwarization \cite{bonati2021intelligence}. These changes provide a range of new and efficient methods for early detection and localization of the source of an attack, allowing the operator to cut off the malicious flow before it spreads through the network.}

\added{Besides the softwarization and disaggregation, the OpenRAN Alliance \cite{bonati2021intelligence} introduces openness and, more importantly, intelligence that reaches the edge of the network. The research in this area mainly focuses on optimizing radio resources and the dynamic reconfiguration of network entities to match the communication requirements \cite{bonati2021intelligence, polese2022understanding}. However, security implications will play a major role in making the architecture a future-proof alternative to traditional \ac{ran} deployments \cite{polese2022understanding}.}

Instead of the common approach of using the \deleted{control mechanisms for performance optimization} \added{core network for \ac{dpi}}, we are carrying out early active attack detection in \deleted{cellular networks} \added{the \ac{ran}}. 
\deleted{We are taking advantage of }
\added{For that purpose, we leverage} the new OpenRAN architecture to collect air interface measurements and \ac{ml} to perform traffic classification (i.e., attack detection). This allows us to detect the attack early on and stop it before it progresses through the network. Our main contributions include: 
\vspace{-0.3em}

\begin{itemize}
    \item We developed a near-\ac{rt} \ac{ric} that allows us to collect measurements from open-source base stations (i.e., \emph{srsRAN}) and dynamically adjust their configuration;
    \item We identify the most important features on the physical and \ac{mac} layer for detecting different types of \ac{dos} attacks;
    \item We designed a \ac{ml} model that allows us to classify different types of \ac{dos} attacks with high accuracy based on the air interface measurements collected by the near-\ac{rt} \ac{ric} (i.e., physical and \ac{mac} layer measurements). 
\end{itemize}

\section{Architecture / Framework}\label{sec:arch}

\begin{figure}[t] 
\centering
\includegraphics[scale=0.9]{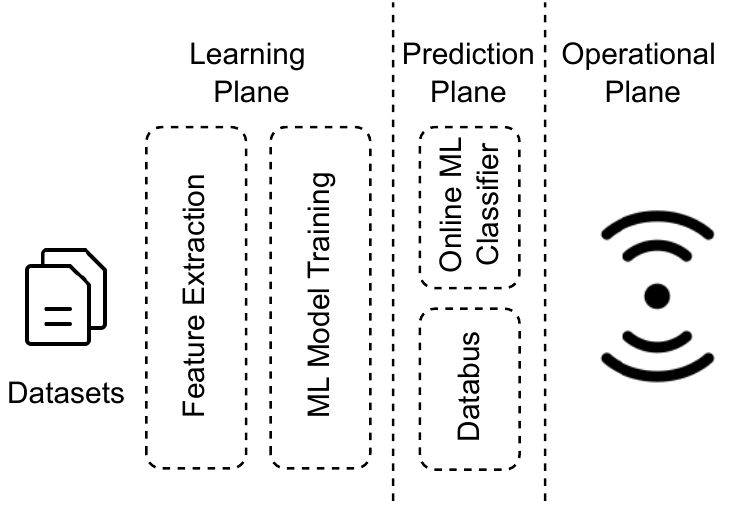}
\caption{Data collection and ML model training and inference framework.}
\label{architecture}
\vspace{-0.5em}
\end{figure}
\newcommand{\bgcell}[2][c]{%
  \textbf{\begin{tabular}[#1]{@{}c@{}}#2\end{tabular}}}
 \newcommand{\cell}[2][l]{%
  \begin{tabular}[#1]{@{}l@{}}#2\end{tabular}}
\begin{table}[t]
\centering{
\caption{Physical and MAC layer features used in our experiments.\noted{BRUNO: I reduced Table I to fit in 1 column, space reduced was not representative. If you (Merim) prefer we can leave it in 2 columns.}}
\begin{tabular}{|c|c|c|l|l|}
\hline
\textbf{Layer} & \bgcell{Down\\link} & \bgcell{Up\\link} & \textbf{Feature} & \textbf{Short Description}                           \\ \hline
\multirow{4}{*}{PHY}  & \checkmark        &             & cqi          & Channel Quality Indicator                            \\ \cline{2-5} 
 & \checkmark        & \checkmark  & mcs          & \cell{Modulation and Coding \\Scheme}                         \\ \cline{2-5} 
 &                  & \checkmark  & \cell{pusch \\sinr}   & \cell{Noise ration interference \\ in the Physical Uplink \\ Shared Channel}   \\ \cline{2-5} 
 &                  & \checkmark  & \cell{pucch \\sinr}   & \cell{Noise ration interference \\ in the Physical Uplink \\Control Channel}  \\ \hline

\multirow{3}{*}{MAC} &\checkmark        & \checkmark  & brate        & Bitrate (bits/unit measure)                          \\ \cline{2-5} 
 & \checkmark        & \checkmark  & pkt ok           & \cell{Number of packets \\ successfully sent}                  \\ \cline{2-5} 
 &\checkmark        & \checkmark  & pkt nok          & Number of packets dropped                            \\ \hline
\end{tabular}
\label{tab:feature}
\vspace{-1em}
}
\end{table}

Our traffic classification framework was built around two main constraints: (1) latency requirement imposed by the near-\ac{rt} \ac{ric}; and (2) low resolution of the features from the physical and \ac{mac} layers needed to describe the network traffic. Both constraints limit the choice of the \ac{ml} classifiers, i.e., the goal is to train a model that can accurately classify the traffic within the defined latency requirements based on a limited set of features available in the physical and \ac{mac} layers. We have developed our own control plane (i.e., near-\ac{rt} \ac{ric}) that allows us to collect measurements and to control the \emph{srsRAN} virtual \ac{bs}. The framework is implemented as an \textit{xApp} running on the near-\ac{rt} \ac{ric}. Fig.~\ref{architecture} shows the required steps to deploy an online \ac{ml} classifier within the proposed architecture.




The first step in the process is to identify the frequency of the measurement collection and the type of measurements needed to perform the traffic classification. These steps are implemented in the \LEARNING. Additionally, the \LEARNING is responsible for storing the data from the \OPERATION. The training of the initial \ac{ml} model is done offline once enough data is collected from the \OPERATION. The first step in the training process is feature extraction. The \emph{Feature Extractor} component identifies the useful measurements and performs the data pre-processing before the training starts. Then, the \emph{ML Model Training} component trains the new model on the same time scale that was chosen for the measurement collection. The \emph{ML Model Training} component also ensures that the new model conforms to the processing requirements set by the near-\ac{rt} \ac{ric}. Once the model is trained, and the target accuracy is achieved, the model can be deployed on the \PREDICTION. 


The \PREDICTION performs multiple tasks: (i) pulls measurements from the \OPERATION; (ii) shares the collected measurements with the \LEARNING for further model refinement; (iii) feeds data to the \emph{Online ML Classifier} for traffic classification; and (iv) depending on the traffic class sends commands to the \ac{bs} in order to continue or terminate the service. The communication between the various components in our architecture is done through the \emph{Databus} of our control plane. The \emph{Databus} is implemented as a \emph{ZMQ} broker\footnote{\url{https://zeromq.org}}.

    
The \OPERATION consists of the \emph{srsRAN} \ac{bs}, the core network and our custom built \ac{ric} agent. The \ac{ric} agent allows us to share data and read commands with/from the \emph{Databus} of the near-\ac{rt} \ac{ric}. Once the \ac{ue} connects to the \ac{bs}, the \ac{ric} agent starts collecting measurements and sending them as asynchronous messages via the \emph{Databus} to the \PREDICTION. The \PREDICTION performs the traffic classification and, if needed, sends control messages to the \ac{ric} agent. The \ac{ric} agent interprets these control messages and controls the operation of the \ac{bs} accordingly (e.g., forward, prioritize, or drop packets from the \ac{ue}). 






As highlighted earlier, only the initial model is trained offline. The \mbox{\LEARNING} continuously collects data during the prediction phase and refines the initial model. In terms of \ac{ml}, attacks are rare events, and as such they are hard to predict. Therefore, as we will present in Section~\ref{sec:result}, besides inference time, we mainly rely on the $F1$-Score to express the model performance. As a reminder the $F1$-Score is computed as:
\vspace{-1em}
\begin{align}\label{equ:f1}
F1 = \frac{2 R P}{R+P}
\end{align}
where $R$ represents recall and $P$ represents precision. This allows us to ensure that the model is penalised equally for being biased towards positive predictions as well as missing predictions. Once the $F1$-Score improves significantly, the new model gets deployed to the \mbox{\PREDICTION}.
\noted{BRUNO: It's a characteristic previewed in the framework, however didn't implemented, we could remove this paragraph.}

\vspace{-0.5em}
\subsection{Feature Extraction}\label{sec:result:feature}

\deleted{As highlighted in the previous section, all components in our architecture communicate through our control plane's \emph{Databus}.} \added{The traditional approach to identify malicious data flows is to extract features from packet headers. In cellular networks, this is usually performed in the core network. However, our goal is to identify and stop these data flows at the source, i.e., the edge of the network. This introduces additional challenges related to the features being less expressive (i.e., measurements collected from the physical and MAC layers don't contain packet headers).} During the initial \ac{ml} model training phase, the data collected from the \ac{bs} gets labeled, and the feature extraction process starts. Table~\ref{tab:feature} describes the features used in the classification process. All presented features are used to describe the quality of the signal and the traffic volume. 
\deleted{Please notice that the features in Table~\mbox{\ref{tab:feature}} are measurements collected from the physical and \mbox{\ac{mac}} layer.}

\noted{BRUNO: We could try to reduce a little this two paragraphs.}

The channel quality is measured by the \ac{ue} and reported back to the \ac{bs} as the \ac{cqi}. The \ac{cqi} index is a scalar value from $0$ to $15$ measured on the physical layer. The \ac{mcs} is also a scalar that is determined based on the reported \ac{cqi}. However, the mapping also depends on the amount of information that is being sent between the \mbox{\ac{ue}} and \mbox{\ac{bs}}. The range of values for the \emph{\ac{mcs}} is between $0$ and $28$, meaning that there is no direct mapping between the \emph{\ac{cqi}} and \emph{\ac{mcs}}. The \ac{pusch} \ac{sinr} and \ac{pucch} \ac{sinr} metrics indicate how much the desired signal is stronger compared to the noise and interference in the two channels, i.e., \ac{pusch} and \ac{pucch}. These metrics provide information about the relationship between the channel conditions and the achieved throughput.




\deleted{Unlike the physical layer measurements mentioned above, the \mbox{\ac{mac}} layer provides statistics on the amount of data and packets exchanged between the \mbox{\ac{ue}} and \mbox{\ac{bs}}. Measurements like \mbox{\emph{brate}}, \mbox{\emph{pkts ok}} and \mbox{\emph{pkts nok}} provide information about the data rate in bits, the number of delivered and dropped packets, respectively. These measurements provide information about the traffic patterns of the services exchanging information through the network.}

\added{
Unlike the physical layer measurements mentioned above, the \ac{mac} layer statistics contain the amount of data and packets exchanged between the \ac{ue} and \ac{bs}, providing information about the traffic patterns of the services exchanging information through the network. These measurements include data rate (\emph{brate}), the number of delivered packets (\emph{pkts ok}) and the number of dropped packets (\emph{pkts nok}).}

\vspace{-0.5em}
\subsection{ML Model Training}\label{sec:arch:model_training}

\noted{BRUNO: Just removing some references}

Once a large enough dataset is collected and the features extracted, we start the \ac{ml} model training phase. Unlike the work presented in \cite{xavier2022tnsm, barradasflowlens, lee2020switchtree}, in which the authors focus on traffic classification with limited computational resources, we do not struggle with those limitations. However, unlike the models presented in those papers, our model does not have access to the intrinsic (i.e., packet headers) and extrinsic features (i.e., inter-packet arrival time). It restricts us to works under the limitations highlighted in Section \ref{sec:arch} (i.e., low latency and low resolution of the available features). 


The near-\mbox{\ac{rt}} \mbox{\ac{ric}} imposes a $1s$ latency threshold. In other words, the overall latency, including measurement collection, network delay between the \ac{bs} and the controller, and the processing delay on the \emph{databus} and the \emph{xApp} has to be lower than $1s$. Please notice that the latency requirement is for the whole control loop, i.e., both ways: measurements traveling from the \ac{bs} to the \ac{ric} and control messages traveling in the other direction. Even though the scope of the paper does not include a detailed analysis of best traffic classification models, in Section~\ref{sec:result}, we included an analysis of the accuracy and performance of the most popular \ac{ml} models that provide a suitable choice for the online classification under the mentioned restrictions.

\vspace{-0.5em}
\subsection{Online ML Classifier}\label{sec:arch:online_classifier}
The traffic classification starts once the \emph{\ac{ml} Model Training} Component converges to a state which achieves the required accuracy within the defined latency bounds. The \emph{Online \ac{ml} Classifier} pulls measurements from the \emph{Databus} and performs the traffic classification for each \ac{ue} connected to the \ac{bs}. Once the traffic class is identified, a command is being sent to the \ac{bs} based on predefined policies (e.g., forward, prioritize, or drop packets from the \ac{ue}). Additionally, it shares the measurements with the \mbox{\LEARNING} for further model refinement.



\section{Proposed Approach}\label{sec:proto}
This section describes the implementation decisions and the interaction between the control components with the \emph{srsRAN} architecture (Section~\ref{sec:approach:setup}). Sections~\ref{sec:approach:class} and \ref{sec:approach:dataset} provide details about the traffic categories and the experimental setup.



\begin{figure}[] 
\centering
\includegraphics[scale=0.26]{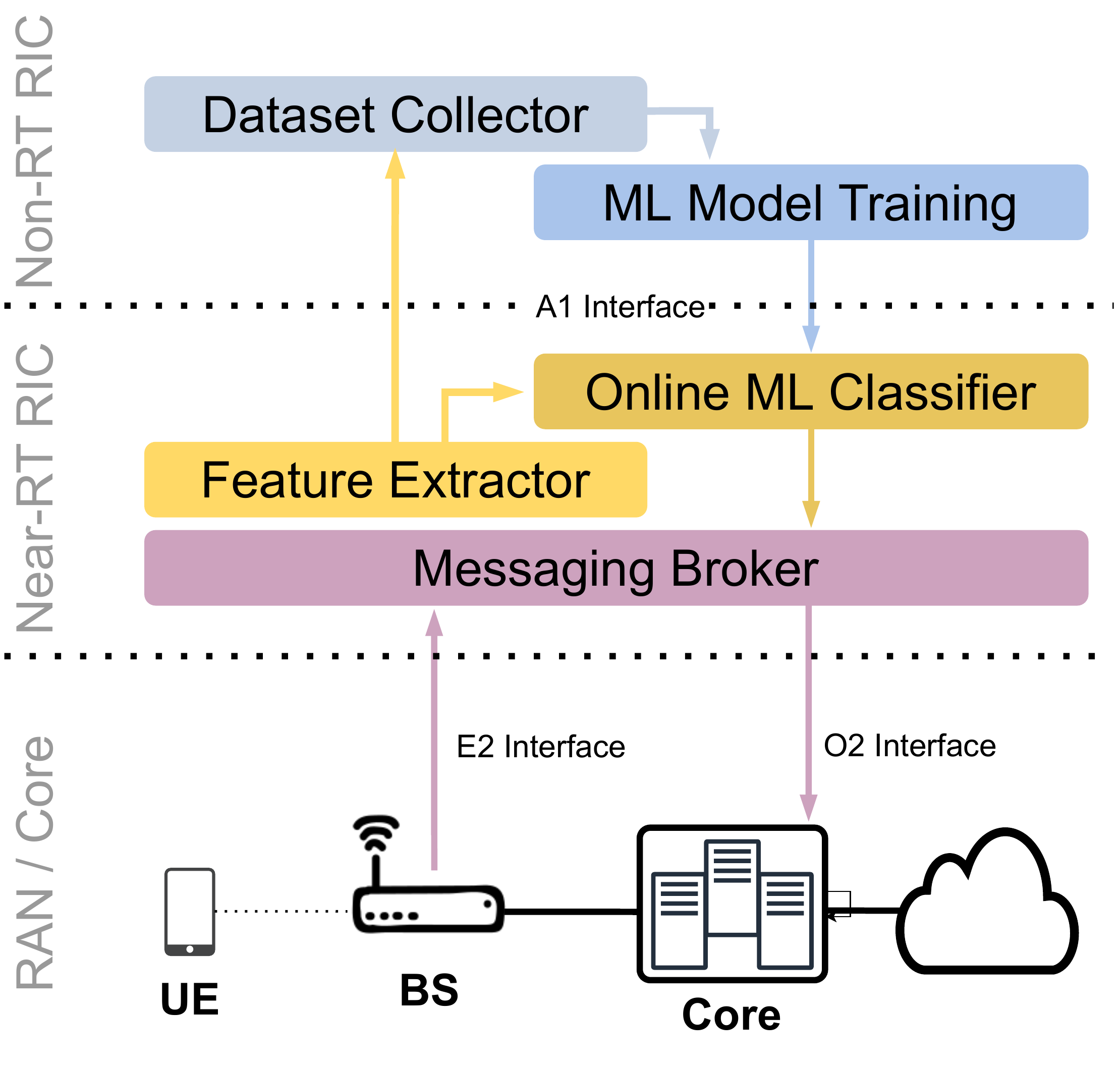}
\caption{Experimental setup architecture describing the interfaces and communication between all components.}
\label{proto:architecture}
\vspace{-1.5em}
\end{figure}

\vspace{-0.5em}
\subsection{Experimental Setup}\label{sec:approach:setup}




As shown in Fig.~\ref{proto:architecture}, our experimental setup consists of three main components: (1) \ac{ran} and core network; (2) near-\ac{rt} \ac{ric}; and (3) non-\ac{rt} \ac{ric}. The system is implemented in the OpenIreland research testbed infrastructure \cite{OpenIreland}.


\subsubsection{\textbf{RAN and Core Network}} \label{setup:ran}
For the purpose of our experiments we used Ettus Research \acp{usrp} and the \emph{srsRAN} open-source 4G and 5G software radio library. The \emph{srsRAN} software was modified, and we developed our own library, i.e., \emph{\ac{ric} Agent}, that is compiled with the \emph{srsRAN} software allowing us to share measurements with the \emph{Databus} in the control plane and read/execute commands coming from the \emph{xApps}. 

Our experimental network consists of $3$ \acp{usrp} - $2$ acting as \acp{ue} and $1$ acting as a \ac{bs}. These \acp{usrp} are connected to \acp{vm} running on a Dell PowerEdge R440 server. One additional \ac{vm} acts as the core network.


\subsubsection{\textbf{Near-RT RIC}} \label{setup:near-RT}
We developed a near-\ac{rt} \ac{ric}, that communicates with the \ac{ran}, allowing us to read measurements from the \acp{bs} and send commands based on the algorithms implemented in the \emph{xApps}. The implementation consists of three main components: (1) the previously mentioned \emph{RIC Agent} that is compiled with the \emph{srsRAN} software; (2) a \emph{Databus} that is implemented as a \emph{ZMQ} broker, allowing us to exchange measurements and commands between the \emph{xApps} and \acp{bs}; and (3) the \emph{xApps} that allow us to implement control mechanisms that take measurements as inputs and make decisions based on optimization algorithms, \ac{ml} models or predefined policies. 

\subsubsection{\textbf{Non-RT RIC}} \label{setup:non-RT}
Unlike the near-\ac{rt} \ac{ric}, we did not develop a non-\ac{rt} \ac{ric}. For the purpose of our experiments the non-\ac{rt} aspects were performed through offline learning and hyperparameter tuning. Once the accuracy and classification delay of the \ac{ml} model fell within the requirements set by the near-\ac{rt} \ac{ric}, we deployed the model as an \emph{xApp}. 

\added{A closer look at Fig.~\ref{proto:architecture} reveals that the total delay of the control loop $T_d$ depends on the round trip transmission delay (i.e., network delay) between all nodes hosting different parts of the architecture $t_n$, the processing time at the \emph{Databus} $\delta_d$, and the inference time on the near-\ac{rt} \ac{ric} $\delta_i$. This can be expressed with Equation~\ref{equ:control_loop_delay}: }
\vspace{-0.6em}
\begin{align}\label{equ:control_loop_delay}
\added{T_d = t_n + 2\delta_d + \delta_i}
\end{align}

\vspace{-0.3em}

The round trip network delay $t_n$ can be expressed as:
\vspace{-0.4em}
\begin{align}\label{equ:network_delay}
\added{t_n = 2\delta_{bd} + 2\delta_{dr}}
\end{align}
\added{where $\delta_{bd}$ is the transmission delay between the \ac{bs} and the \emph{Databus}, and $\delta_{dr}$ is the tranmsision delay between the \emph{Databus} and the near-\ac{rt} \ac{ric}.}

To understand the control loop delay margins required to satisfy the near-\ac{rt} control requirements, we provide delays for all involved components. The delay of the \emph{Databus} is $\delta_d=45\mu s$ and the round trip network delay of our setup is $t_n=670\mu s$. This results in a total control loop delay $T_d=760\mu s + \delta_i$. Please notice that our round trip network delay is very low (i.e., $< 1ms$) due to the setup being physically collocated. In Section~\ref{sec:result}, we will further investigate the inference time $\delta_i$ to compute the margin for the overall round trip network delay.



\vspace{-0.2em}
\subsection{Traffic Categories}\label{sec:approach:class}
In order to collect a realistic dataset, we used well known tools to generate traffic in our network. We considered two traffic categories, i.e. \emph{Benign} and \emph{Attack}. Each traffic category was represented by the following network applications: 
\begin{itemize}
    \item \emph{Benign}: \emph{Web Browsing} and \emph{\ac{voip}};
    \item \emph{Attack}: \emph{\ac{ddos} Ripper}, \emph{\ac{dos} Hulk} and \emph{Slowloris}. 
\end{itemize}


\subsubsection{\textbf{Web Browsing}}
Web Browsing is characterized by the variable amount of data transmitted to the web server and received by the client. Since the majority of internet applications use \ac{tcp}, in order to generate \emph{Web Browsing} traffic, we access the most visited websites online\footnote{\url{https://en.wikipedia.org/wiki/List_of_most_visited_websites}}, and navigate through them randomly. 





\subsubsection{\textbf{\ac{voip}}} 
\ac{voip} traffic has strict \ac{qos} requirements. It relies on \ac{udp} due to the lower tolerance to delays than packet drops. A typical voice call requires between $20$ kbps and $170$ kbps guaranteed priority bandwidth. To simulate the voice application, we rely on \emph{SIPp}\footnote{\url{http://sipp.sourceforge.net/}}, which is an open-source traffic generator widely used to test \ac{voip} services in the real environment.


\subsubsection{\textbf{DDoS Ripper}} 
\emph{\ac{ddos} Ripper} explores the webserver vulnerability flooding it with requests that result in cutting off targets or surrounding infrastructure. It opens as many connections with the server as possible and keeps them alive, sending a high volume of trash headers. This packet flooding causes an unexpected traffic jam preventing legitimate traffic from reaching its destination. Our code is based on the \emph{PyPI \ac{ddos} Ripper}\footnote{\url{https://pypi.org/project/ddos-ripper/}}.

\subsubsection{\textbf{DoS Hulk}} 
This attack relies on obfuscating the traffic as a benign request. However, it sends an unbearable load of \ac{tcp} SYN messages and a flood of HTTP GET requests through a multi-thread agent. Therefore, this attack requires a considerable throughput. Our \ac{dos} Hulk script relies on \emph{PyPI}\footnote{\url{https://pypi.org/project/hulk/}}.


\subsubsection{\textbf{Slowloris}} 
\emph{Slowloris} is a type of attack that opens multiple connections to the webserver and keeps it open as long as possible. It requires a very low bandwidth to periodically send subsequent HTTP headers for each opened connection. This traffic pattern resembles benign communication by using the legitimate packet headers to keep connections alive. We based our  \emph{Slowloris} Python script on the \emph{PySloloris}\footnote{\url{https://pypi.org/project/pyslowloris/}} development.


\added{To ensure that the network traffic behaves according to real world patterns, our approach is based on the techniques recommended by \cite{sharafaldin2018toward}.}



\vspace{-0.5em}
\subsection{Dataset Collection}\label{sec:approach:dataset}



In order to train and validate our traffic classification \emph{xApp}, we collected two datasets. The first was collected by running the traffic generator scripts on one \ac{ue}, randomly switching between the different types of traffic patterns. These experiments require only $2$ \acp{usrp} were used, i.e., one \ac{ue} and one \ac{bs}. The dataset was collected by the near-\ac{rt} \ac{ric}.



The second dataset was collected by randomly switching between different types of traffic on two different \acp{ue}. This setup involves $3$ \acp{usrp}, i.e., two \acp{ue} and one \ac{bs}. Again, the near-\ac{rt} \ac{ric} was used to collect the measurements on the \ac{bs}. In this case, the dataset includes the \ac{bs} measurements, the traffic label and the label predicted by the \emph{Online Classifier}.


Table~\ref{tab:datsets} shows the number of samples collected for the abovementioned scenarios. Section~\ref{sec:result} provides details about fitting and evaluating the \ac{ml} models in the training phase, and details about the performance of the \emph{Online \ac{ml} Classifier}. 
\vspace{-0.5em}


\section{Experimental Results}\label{sec:result}

In this section, we investigate the possibility to perform attack detection based on the limited features available in the \ac{ran}. The performance of the \emph{xApps} is evaluated based on the classification accuracy and the inference delay for the most popular \ac{ml} algorithms used for traffic classification.  

\begin{table}[t]
\centering{
\caption{Number of samples per traffic class for both datasets.}
\begin{tabular}{|p{3cm}|l|l|}
\hline
\textbf{Class} & \textbf{1 UE} & \textbf{2 UEs} \\ \hline
Web Browsing            & 32,135         & 300,798        \\ \hline
VoIP           & 69,802         & 121,975       \\ \hline
DDoS Ripper    & 33,083         & 86,246        \\ \hline
DoS Hulk       & 34,980         & 78,339        \\ \hline
Slowloris      & 26,078         & 492,145       \\ \hline \hline
Benign         & 101,937        & 422,773       \\ \hline
Atack          & 94,141         & 656,730       \\ \hline
\end{tabular}
\label{tab:datsets}
\vspace{-1em}
}
\end{table}

\begin{table}[t]
\centering{
\caption{Accuracy, Fitting and Inference time for all tested models.}
\begin{tabular}{|l|c|c|c|}
\hline
\textbf{}     & \textbf{Accuracy} & \textbf{\begin{tabular}[c]{@{}c@{}}Fitting \\ (sec) \end{tabular}} & \textbf{\begin{tabular}[c]{@{}c@{}}Inference \\ (ms)\end{tabular}} \\ \hline
SVM           & 0.64            & 1,952.220    & 3.02   \\
$k$-NN        & 0.90            & 0.793        & 2.10      \\ 
Decision Tree & 0.93            & 0.627        & 1.25      \\ 
Random Forest & 0.95            & 0.783        & 2.86      \\
AdaBoost      & 0.87            & 19.615       & 3.05      \\ 
MLP           & 0.56            & 15.720       & 1.29      \\ \hline
\end{tabular}
\label{tab:inference_time}
\vspace{-2em}
}
\end{table}

As previously highlighted, the \emph{xApp} has to identify the traffic class within the acceptable delay constraints defined by the near-\ac{rt} \ac{ric}, while still maintaining a high level of accuracy. Our evaluation considers six different \ac{ml} classifiers, i.e., \ac{svm}, \ac{knn}, Decision Tree, Random Forest, \ac{adaboost} and \ac{mlp}. We use the dataset collected by connecting one UE to the \ac{bs} for training. Therefore, the first column of Table~\ref{tab:datsets} shows the number of samples used for the training of the classifiers for each traffic class. We use the second dataset (i.e., second column in Table~\ref{tab:datsets}) for validation and testing purposes.
\vspace{-1.3em}


\hfill
\begin{table}[t]
\centering{
\caption{Summary of the online classification results.}
\begin{tabular}{p{3.5cm}|l|l|l}
\hline
\hline

\textbf{Application} & \textbf{Precision} & \textbf{Recall} & \textbf{F1-score} \\
\hline
Web Browsing              & 0.93 & 0.96 & 0.94  \\
VoIP             & 0.97	& 0.95	& 0.96   \\
DDoS Ripper      & 0.94	& 0.87	& 0.90  \\ 
DoS Hulk         & 0.87	& 0.95  & 0.91   \\
Slowloris        & 0.97 & 0.95	& 0.96   \\
\hline
Benign           & 0.94 & 0.96 & 0.95 \\
Attack           & 0.98 & 0.96 & 0.97 \\
\hline
\hline
\end{tabular}
\label{tab:class_res}
\vspace{-1em}
}
\end{table}

\begin{figure}[t] 
\centering
\includegraphics[scale=0.47]{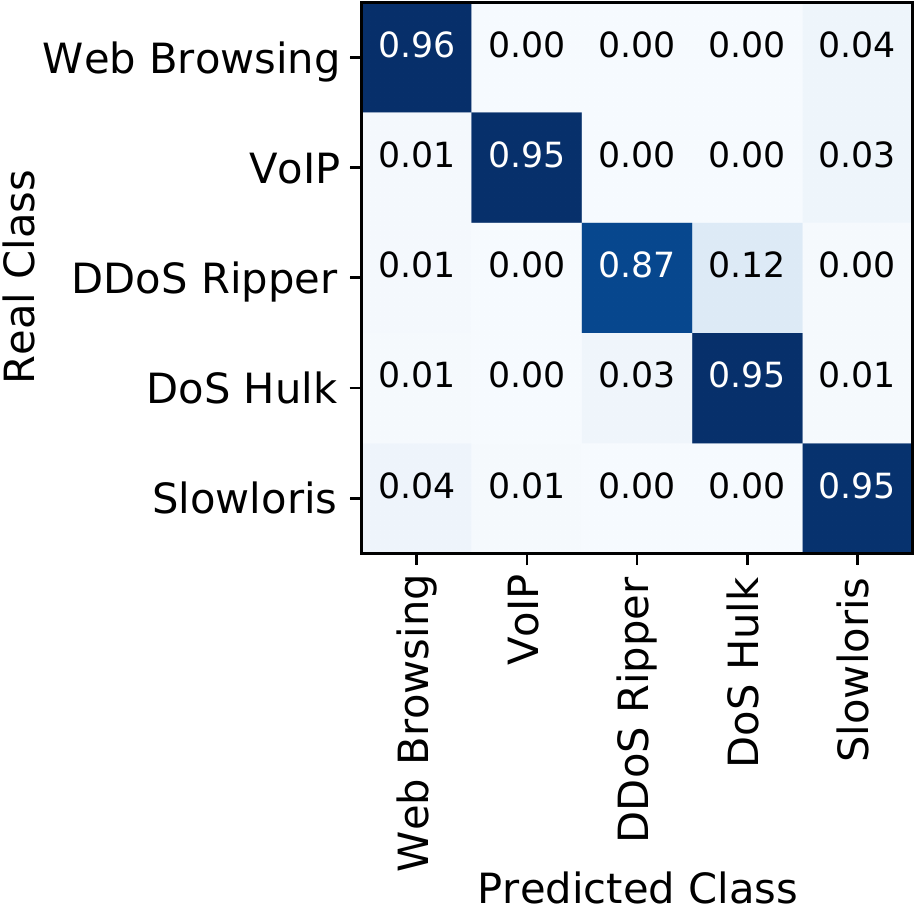}
\includegraphics[scale=0.47]{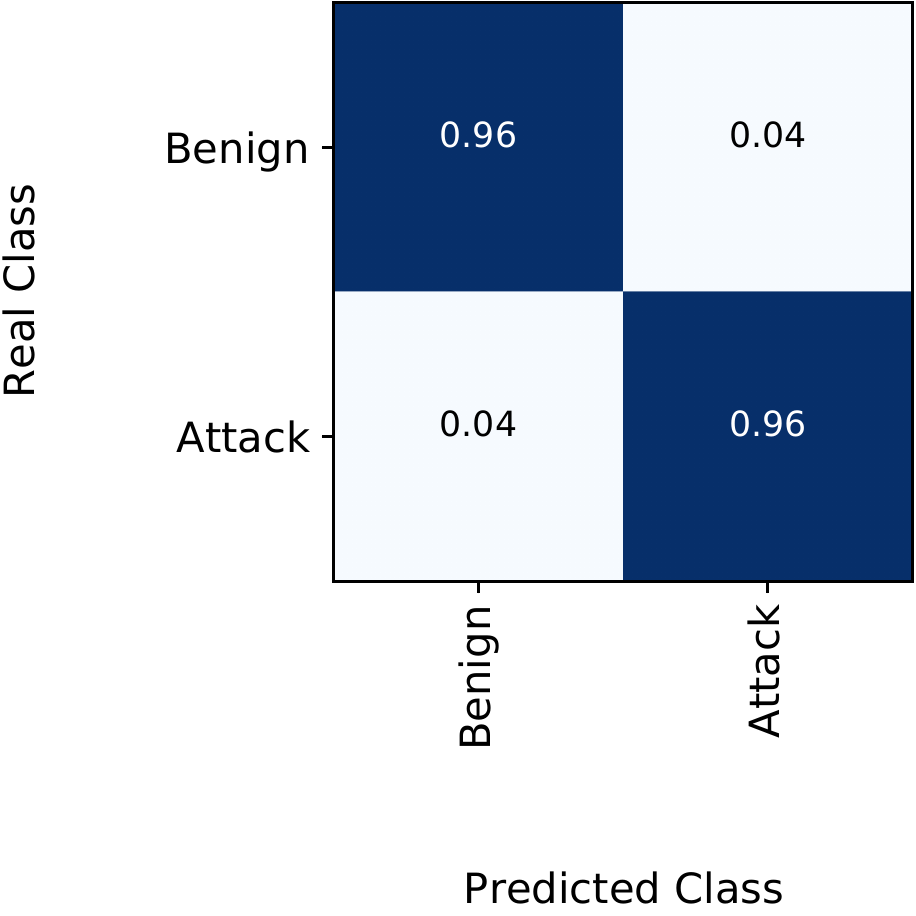}

\caption{Confusion matrices for the individual classes (left), and the binary problem (right).}
\label{fig:conf_matrix}
\vspace{-1em}
\end{figure}

Due to the nature of real network traffic, different traffic classes can exhibit similar properties during short periods of time. Our analysis showed that in such scenarios, traffic classification models based on decision rules have an advantage due to the higher tolerance to outliers. Additionally, they have a lower inference delay - $\delta_i$ due to lower computational requirements. Table~\ref{tab:inference_time} shows the classification accuracy achieved by different \ac{ml} algorithms. The Random Forest and Decision Tree algorithms outperformed the others with an accuracy of $0.95$ and $0.93$, respectively. As shown in Table~\ref{tab:inference_time}, the accuracy of the the other algorithms is considerably lower with the minimum being $0.56$ for \ac{mlp}. 

A closer inspection of Table~\ref{tab:inference_time} reveals that the inference time - $\delta_i$ of all tested classifiers falls within the acceptable bound set by the near-\ac{rt} \ac{ric}. It ranges from $1.25ms$ for the Decision Three to $3.05ms$ for the \ac{adaboost} classifier. Due to the high accuracy discussed above and the fact that the inference time of all tested classifiers is acceptable, the Random Forest is the best choice for our use-case. \added{We argue that the}\deleted{The} Decision Tree classifier would be the second choice considering that its inference time is much lower, with the accuracy being lower by $2\%$. 




Considering the analysis presented above, we deployed the Random Forest classifier as an \emph{xApp} to the near-\ac{rt} \ac{ric}. The chosen model consists of $100$ decision trees with a maximum depth of $15$. The minimum number of samples required to split an internal node is $5$, and the minimum number of samples required to reach a leaf node is $1$. The experiments were executed in our testbed with the setup described in Section~\ref{sec:approach:setup}. Even though the real testbed setup exhibits variations in the channel quality, and therefore achievable throughput, the deployed classifier was able to achieve the same classification accuracy that was achieved during the offline training and testing phase, i.e., $0.95$. 


Table~\ref{tab:class_res} provides details about the precision, recall and $F1$-Score for all traffic classes used in our experiments. These metrics show that the proposed model can predict different traffic classes based on measurements taken from the physical and \ac{mac} layer. Fig.~\ref{fig:conf_matrix} provides further evidence of the network traffic classification quality. For the individual traffic classes the model achieves a $0.93$ $F1$-Score (Fig.~\ref{fig:conf_matrix}, left). 


So far, we focused on identifying the individual traffic classes. However, in many applications, the goal is to distinguish between \emph{Benign} traffic and an \emph{Attack}. The right hand side of Fig.~\ref{fig:conf_matrix} shows the confusion matrix for these two categories. It clearly shows that our classifier correctly identifies $96\%$ of the \emph{Attacks} while achieving a very high $0.96$ $F1$-Score. \added{Due to the lack of comparable solutions in the \ac{ran}, we compare these results to \ac{ids} solutions deployed in the core network. Even though \ac{ids} solutions have an advantage due to the availability of packet headers, our Open-\ac{ran} approach achieves results comparable to the work in \cite{xavier2022tnsm} and \cite{barradasflowlens} in which the authors report 97\% $F1$-Score and 93\% accuracy, respectively. The authors of \cite{xavier2022tnsm} report that their \ac{ids} solution introduces additional $3.7\mu s$ of delay per packet. In contrast, the near-\ac{rt} \ac{ric} does not interfere with the traffic flow, meaning that no additional delay is introduced to the network. }



\begin{figure}[!t] 
\centering
\includegraphics[scale=0.9]{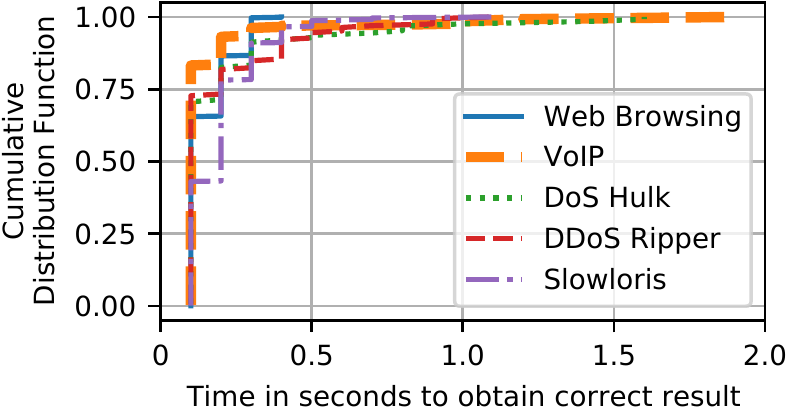}
\caption{Cumulative Distribution Function showing that the network traffic is in a transition state for the first 500ms, after which we can correctly classify all analyzed traffic classes.}
\label{fig:pkts_correct_class}
\vspace{-1.8em}
\end{figure}

Fig.~\ref{fig:pkts_correct_class} shows that after $500ms$, more than $90\%$ of the service executions will be classified correctly. This delay is due to the time needed for the traffic to normalize its behavior, i.e., all analyzed traffic patterns exhibit an initial transition phase in which they can't be correctly classified. Please note that in our experiments we collect measurements every $100ms$. This means that after $5$ measurements and a negligible network - $t_n$ and \emph{Databus} processing delay - $\delta_d$ (see Section~\ref{sec:approach:setup}), we correctly classify more than $90\%$ of the traffic executions. 

As highlighted throughout the paper, real-time traffic classification is important for on-demand resource allocation, dynamic network reconfiguration and, in case of the detection of potential network attacks, it is useful for \acp{ids}. Motivated by the idea of real-time traffic classification, we measured the time needed to correctly label each traffic type. Table~\ref{tab:inference_time} shows that the inference time is $\delta_i=2.86ms$. Once this value is added to the previously computed control loop delay, we get to a total of $T_d=760\mu s+2.86ms=3.62ms$, meaning that there is a high margin for the network delay (i.e., $996ms$) that still allows us to operate within the $1s$ control loop delay as defined by the Open-\ac{ran} Alliance. In terms of detecting network attacks it allows us to terminate the service for the malicious \ac{ue} before any damage is made. Based on the proposed architecture and the availability of measurements, the detection of the attacks can be done at the \ac{du}, while the attack could be stopped at the \ac{cu}. The \ac{bs} can indicate the \emph{\ac{rrc} Connection Release} procedure to the \ac{ue}, which will effectively terminate the service for the malicious user, by moving the \ac{ue} from \emph{RRC\_Connected} to \emph{RRC\_Idle} state.




\vspace{-0.4em}
\section{Conclusions and Future Work}
We developed a near-\ac{rt} \ac{ric} and modified the open-source base station implementation from \emph{srsRAN} to perform early \ac{dos} attack detection. This framework allows us to collect measurements and control the \ac{ran} configuration with delays between $10ms$ and $1s$. In order to comply with these delay requirements, we developed a fast \emph{Databus} with a delay of $45\mu$s and tested the inference delay of various \ac{ml} algorithms.

In our experiments, we also extract the most important air interface measurements/features required to correctly classify the analysed network traffic. It is important to notice that these measurements don't provide details about the packet headers. For the purpose of testing the proposed architecture, we collected a realistic dataset in our testbed. The real setup allowed us to validate the performance of our \ac{ml} model in an environment with varying channel conditions. Our results show that the proposed model generalizes well and it was able to correctly classify the traffic with high accuracy. 

The results indicate that early \ac{dos} attack detection is possible at the edge of the network, allowing us to isolate the malicious users and stop the attack before it does any damage to the rest of the network. Future work on this topic would include the implementation of the Real Time \ac{ric}, which operates on even lower delay requirements (i.e. up to $10ms$). This would also affect the type of \ac{ml} model that could perform the inference. 


\hfill
\hfill
\hfill

\begin{acronym}
  \acro{ddos}[DDoS]{Distributed Denial-of-Service}
  \acro{sdn}[SDN]{Software-Defined Networking}
  \acro{ml}[ML]{Machine Learning}
  \acro{mimo}[MIMO]{Multiple-Input and Multiple-Output}
  \acro{ran}[RAN]{Radio Access Network}
  \acro{rt}[RT]{Real Time}
  \acro{ric}[RIC]{RAN Intelligent Controller}
  \acro{mac}[MAC]{Medium Access Control}
  \acro{bs}[BS]{Base Station}
  \acro{ue}[UE]{User Equipment}
  \acro{cqi}[CQI]{Channel Quality Indicator}
  \acro{mcs}[MCS]{Modulation and Coding Scheme}
  \acro{pusch}[PUSCH]{Physical Uplink Shared Channel}
  \acro{pucch}[PUCCH]{Physical Uplink Control Channel}
  \acro{sinr}[SINR]{Signal-to-Interference-plus-Noise Ratio}
  \acro{usrp}[USRP]{Universal Software Radio Peripheral}
  \acro{vm}[VM]{Virtual Machine}
  \acro{voip}[VoIP]{Voice over IP}
  \acro{dos}[DoS]{Denial-of-Service}
  \acro{tcp}[TCP]{Transmission Control Protocol}
  \acro{qos}[QoS]{Quality of Service}
  \acro{udp}[UDP]{User Datagram Protocol}
  \acro{svm}[SVM]{Support Vector Machine}
  \acro{knn}[k-NN]{k-Nearest Neighbors}
  \acro{adaboost}[AdaBoost]{Adaptive Boosting}
  \acro{mlp}[MLP]{Multilayer Perceptron}
  \acro{ids}[IDS]{Intrusion Detection System}
  \acro{du}[DU]{Distributed Unit}
  \acro{cu}[CU]{Centralized Unit}
  \acro{rrc}[RRC]{Radio Resource Control}
  \acro{dpi}[DPI]{Deep Packet Inspection}
\end{acronym}

\vspace{-1.5em}
\section*{Acknowledgment}
Financial support from Science Foundation Ireland 17/CDA/4760, 18/RI/5721 and 13/RC/2077\_p2 is acknowledged. 
Financial support from Brazilian agencies: CNPq, CAPES, FAPESP/MCTI/CGI.br (PORVIR-5G 20/05182-3, and SAWI 20/05174-0), FAPES (94/2017, 281/2019, 515/2021, 284/2021, 1026/2022). CNPq fellows Dr. Martinello 306225/2020-4.

\vspace{-0.8em}

\bibliographystyle{IEEEtran}
\bibliography{IEEEabrv, bib}

\end{document}
